\documentclass{aa}  
\usepackage{graphics}

\def\kms{km\,s$^{-1}$}

\def\nii{[N{\sc ii}]}  
\def\sii{[S{\sc ii}]}

\def\oiii{[O{\sc iii}]}  
\def\cliii{[Cl{\sc iii}]}  
\def\ha{H$\alpha$}  
\def\hb{H$\beta$}  
  
\def\hb{H$\beta$}

\def\cbeta{{$c_\beta$}}  
  
\usepackage{epsfig}  

\begin{document}              
  
\title{Absolute spectrophotometry of northern compact \\ planetary  
nebulae\thanks{Based on observations obtained at the 2.5m~INT  
telescope of the Isaac Newton Group and the 2.6m~NOT telescope  
operated by NOTSA in the Observatorio del Roque de Los  
Muchachos, and with the NASA/ESA Hubble Space Telescope, obtained at  
the Space Telescope Science Institute, which is operated by AURA for  
NASA under contract NAS5-26555.}}  
  
\subtitle{}  
  
\author{S.A. Wright\inst{1,2} \and R.L.M. Corradi\inst{1}   
\and M. Perinotto\inst{3}}  
  
\offprints{R.L.M. Corradi}  
  
\authorrunning{Wright et al.}  
  
\titlerunning{Absolute spectrophotometry of planetary nebulae}  
  
\institute{  
	 Isaac Newton Group of Telescopes, Ap. de Correos 321,  
	 E-38700 Sta. Cruz de la Palma, Canary Islands, Spain\\e-mail:   
	 rcorradi@ing.iac.es  
	\and
	UCLA, Department of Physics and Astronomy, Los Angeles, CA, 
	90095, USA\\email: saw@astro.ucla.edu
        \and    
	 Dipartimento di Astronomia e Scienza dello Spazio,  
         Universit\'a di Firenze, L.go E. Fermi 5, I-50125 Firenze, Italy,  
         e-mail: mariop@arcetri.astro.it}  
\date{}  
  
\abstract{We present medium-dispersion spectra and narrowband images  
of six northern compact planetary nebulae (PNe): \object{BoBn 1}, \object{DdDm 1},
\object{IC 5117}, \object{M 1-5}, \object{M 1-71}, and \object{NGC 6833}.
From broad-slit spectra, total absolute fluxes  
and equivalent widths were measured for all observable emission lines. 
High signal-to-noise emission line fluxes of \ha, \hb, \oiii, \nii, and HeI
may serve as emission line flux standards for northern hemisphere 
observers. From narrow-slit spectra, we derive systemic radial velocities.
For four PNe, available emission line fluxes were measured with 
sufficient signal-to-noise to probe the physical properties of their
electron densities, temperatures, and chemical abundances. 
\object{BoBn 1} and \object{DdDm 1}, both type IV PNe, have an \hb~flux over
three sigma away from previous measurements. We report the
first abundance measurements of \object{M 1-71}. \object{NGC 6833} measured radial velocity and galactic coordinates suggest that it is associated
with the outer arm or possibly the galactic halo, and its low abundance
([O/H]=1.3x10$^{-4}$) may be indicative of low metallicity within that region.}
 
\maketitle  
  
\keywords{planetary nebulae: general -- techniques: spectroscopic:  
emission-line -- flux: standards: chemical abundances}  
  
\section{Introduction}  
\label{S-intro}  
Measuring absolute emission-line fluxes of planetary nebulae (PNe) 
is performed for a number of reasons.  Primarily, absolute fluxes 
can be related to basic properties of the nebulae and their central 
stars (i.e. masses and luminosities, once distances are 
independently known), and can be used to determine nebular 
physico-chemical properties. If total flux of a PNe is accurately measured, 
monitoring over decades can also reveal possible luminosity variations, which  
in turn are related to fast evolution of the central star during the 
PNe phase, especially for the largest core masses.
 
In addition, absolute emission-line fluxes of PNe may also be used 
as standard calibrations for narrow band photometry. In fact, while 
highly precise broad band photometric (Johnson \& Harris \cite{Joh54}; 
Landolt \cite{Lan92}) and spectrophotometric (e.g., Stone \& Baldwin 
\cite{Sto83}; Massey et al \cite{Mas88}; Oke \cite{Oke90}) standard 
stars are presently available, there is still a deficiency of 
emission-line flux standards, which is a requisite for high precision 
photometric study of ionized nebulae (PNe, HII regions, cataclysmic 
variables, supernova remnants, etc.). For narrow band photometry, 
broad band standards observed through a specific filter may not be 
easily related to the flux of an emission-line source. The reason is 
that the transmission curve of the filter is often not accurately 
known, as it depends on many parameters such as temperature during 
observing, tilt, focal ratio, axial angle within the instrument, and 
filter age. In addition, several emission lines coincide with 
absorption lines present in continuum standards (e.g. Balmer series), 
whose exact contribution in narrow band filters is difficult to estimate. 
 
In the past, a number of emission-line standards were derived with use 
of photon counting detectors coupled with interference filters (Liller 
\cite{Lil55}; O'Dell \cite{Odl63}), but they have relatively large 
errors.  In the northern hemisphere, standards primarily for 
\hb\ fluxes were derived by calibrating compact PNe with 
spectrophotometric stars, by means of photoelectric scanning 
observations (e.g. Liller \& Aller \cite{Lil54}, Capriotti \& Daub 
\cite{Cap60}, Miller \& Mathews \cite{Mil72}, Barker \cite{Bar78}). 
Recently, to improve on similar conventional techniques in the 
southern hemisphere, Dopita \& Hua (\cite{Dop97}) derived a set of 
southern emission-line standards by means of broad-slit absolute 
spectrophotometry of a sample of compact PNe.  
Similar emission-line standards are completely missing in the northern 
hemisphere.  
 
Along these lines, we have selected a small sample of (supposedly) 
compact PNe observable from La Palma, and attempted to measure their 
integrated emission-line fluxes.  To make full use of the wide 
spectral coverage, we have measured absolute fluxes and equivalent 
widths for all emission lines present in our observations to 
investigate further the physical and chemical properties of these PNe. 
In Section~\ref{S-Observ/Reduc} we explain the spectroscopic and 
narrowband imaging observational details.  Section~\ref{S-morphology} 
discusses the morphology and sizes of these PNe, and 
Section~\ref{S-analysis} presents the analysis and results for fluxes, 
equivalent widths, radial velocities and chemical 
abundances. Discussion and comparison of our results to literature are 
made in Section~\ref{S-conclusions}. 
  
%
  
\begin{figure}   
\resizebox{8.9cm}{!}{\includegraphics{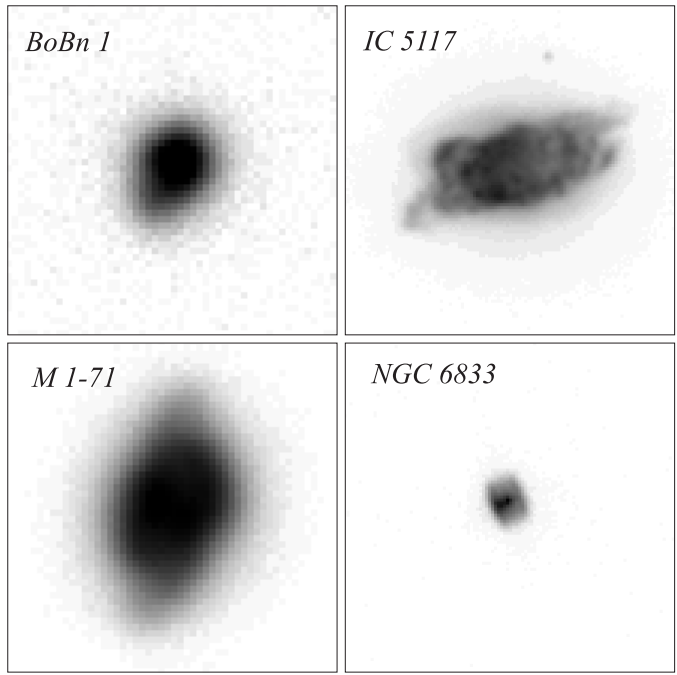}}   
\caption{\nii\ images of four PNe discussed in this paper.  North is 
at the top, East to the left. The field of view is 
10\arcsec$\times$10\arcsec\ for the NOT images of \object{BoBn 1}  
and \object{M 1-71}, and 5\arcsec$\times$5\arcsec\ for the HST images of 
\object{IC~5117} and \object{NGC~6833}.} 
\label{F-images} 
\end{figure} 
 
\section{Observations and reduction}   
\label{S-Observ/Reduc}  
  
\subsection{Spectroscopy}  
 
Spectroscopic observations were obtained at the 2.5m Isaac Newton 
Telescope (INT) of the Isaac Newton Group of Telescopes, Roque de los 
Muchachos, La Palma, using the Intermediate Dispersion Spectrograph 
(IDS). Each observation night was conducted under photometric 
conditions with seeing mostly between 1 and 1.5 arcsec, and 
only rarely as high as 2.5 arcsec.  Blue spectra, covering the range 
370-570~nm, were obtained during three nights on May 
10, July 11, and August 12, 2001 with the EEV10 CCD.  Red spectra, 
covering the range 520-710~nm, were obtained during two nights on 
August 27 and September 4, 2001 with the Tek5 CCD. Both red and blue 
portions were measured with a 632 lines mm$^{-1}$ grating and 235~mm 
camera yielding 0.06~nm pixel$^{-1}$ for the blue spectra and 0.16~nm 
pixel$^{-1}$ for the red spectra. 
  
For each object, a series of exposures were taken with increasing   
exposure times from 60 seconds to 600 seconds, to achieve both  
bright and faint emission lines without saturation. The list of objects  
observed may be found in Table~\ref{obstable}.  
 
Each exposure was observed with the spectrograph slit 
aligned to the parallactic angle to avoid loss of light due to 
atmospheric differential refraction. We used a broad slit with a width 
of 7\arcsec to ensure as much light as possible from the object 
passed through the slit. Further discussion of slit loss may be found 
in Section~4. We used a narrower slit width of 1\farcs5 for radial 
velocity measurements, and observed CuNe arc lamps either before or 
after each observation. 
  
\begin{table*}  
\caption[]{Observational details of target PNe. Diameters are computed.}  
\begin{flushleft}  
\centering  
\begin{tabular}{lccccc}   
\hline \hline  
Object  & PN G & \multicolumn{2}{c}{R.A.\,  (J2000) \,  Dec} & Diameter &  V$_\odot$ \\               
Name    & Number      &        &     &                     [\arcsec]& [\kms]    \\    
\hline                                         
\object{BoBn 1}   & 108.4$-$76.1 & 00 37 18.0 & $-$13 42 12 & 1.5$\times$2.2 & $+174\pm10$\\          
\object{DdDm 1}	  & 061.9$+$41.3 & 16 40 19.4 & $+$38 42 03 & 1.4            & $-317\pm13$\\   
\object{IC 5117}  & 089.8$-$05.1 & 21 32 31.0 & $+$44 35 48 & 1.6$\times$3.5 & $-47\pm6$  \\   
\object{M 1-5}    & 184.0$-$02.1 & 05 46 50.2 & $+$24 22 02 & 2.3$\times$2.8 & $+28\pm7$\\  
\object{M 1-71}	  & 055.5$-$00.5 & 19 36 26.5 & $+$19 42 20 & 3.7$\times$6.0 & $+42\pm8$  \\  
\object{NGC 6833} & 082.5$+$11.3 & 19 49 46.6 & $+$48 57 40 & 0.5$\times$0.6 & $-110\pm2$ \\  
\hline  
\end{tabular}  
\end{flushleft}  
\label{obstable}  
\end{table*}  
  
IRAF software tasks in {\em twodspec} and {\em image} packages  
were used as follows for data reduction. The CCD's overscan region was  
used for bias level subtraction in each image. For each night, a set  
of either dome flats or sky flats were used to remove pixel-to-pixel  
variation, and to eliminate cosmetic defects associated with the  
instrument. Row-by-row wavelength calibration was performed on the  
two-dimensional image, to account for any variations in dispersion in  
the spatial direction.   

The task {\em apall} was used to define an extraction 
aperture large enough to include all of the emission-lines spatially, 
trace the object along the dispersions direction, subtract the 
background, and extract the object. Background subtraction was conducted 
at each wavelength from the median of the background regions specified  
individually for each observation.    
 
Flux calibration was carried out on all PNe with the general set of  
spectrophotometric standards (Stone \& Baldwin \cite{Sto83}, Baldwin  
\& Stone \cite{Bal84}, Massey et al. \cite{Mas88}, Oke  
\cite{Oke90}). Standards were observed at varying times and airmass  
throughout the night to derive a nightly extinction. The  
tasks {\em standards}, {\em sensitivity}, and {\em calibrate}, in  
order, were used for integrating the standard star fluxes over the  
calibrated bandpasses, defining a sensitivity function and extinction  
values, and flux calibrating each PNe.  
 
Spectra taken with the narrow slit were used to determine systematic  
radial velocities for five of the target PNe (\object{BoBn 1}, \object{DdDm 1},
 \object{IC 5117}, \object{M 1-71}, and \object{NGC 6833}). For \object{M 1-5},
additional spectroscopic observations on September 24, 2004 were obtained at the 4.2m 
William Herschel Telescope with the ISIS red arm instrument. A spectral 
range of 607-713 nm was measured using the 500 mm camera and 1200 lines
mm$^{-1}$, yielding a scale of 0.23 nm pixel$^{-1}$. A slit width 
of 0.\arcsec5 was used. Radial velocites for all
PNe were computed using the average wavelength shift of spectral lines
relative to their rest wavelengths, with the IRAF task {\em rvidlines}. 
 
%

\subsection{Images}  
  
Compactness of target PNe is a basic requirement to ensure that 
the total flux of the nebula is included in our broad slit 
spectroscopic observations. Target PNe were chosen to have a diameter 
smaller than 4\arcsec\ in Acker et al. (\cite{SESO}; hereafter A92). Sizes of the 
targets were further checked using narrowband images taken at the INT 
and 2.6~Nordic Optical Telescope (NOT) of La Palma, and using HST 
archive images. 
  
Images from NOT were obtained on September 9, 19, 20, and October 
21, 1997, using the ALFOSC instrument Loral 2k$\times$2k CCD, with a 
spatial scale of 0\farcs19 per pixel. Narrowband filters were used to 
isolate the emission \nii\ at 6583\AA\ from the nearby hydrogen 
H$\alpha$.  For \object{BoBn 1}, an \oiii\ 5007\rm {\AA} image was 
also taken. Exposure times varied between 30 sec and 10~min, depending 
on the brightness of the nebula in each emission line and owing to the 
presence of thin clouds on some nights. The seeing varied between 
0\farcs7 and 1\farcs0. 
 
A 60~sec image of M~1-5 was obtained with the prime-focus Wide Field 
Camera (WFC, pixel scale 0$''$.33) at the INT, on January 29, 
2004. The filter used included both the H$\alpha$ line and [NII] 
doublet. Seeing was 1$''$.4. 
 
Narrowband images of \object{IC~5117} and \object{NGC~6833} were 
retrieved from the HST public archive. They were taken with the PC CCD 
of the WFPC2 camera for programmes n.~6943 and n.~8307 on 1997 and 
1999, respectively, using several narrowband filters including \ha, 
\nii, and \oiii. Exposure times range from 3~sec to 10~min. 
 
\section{Morphology and sizes} 
\label {S-morphology} 
 
As is common for PNe, the largest extension of the nebulae shows up in 
the low excitation line of \nii, as compared to e.g. \ha\ and \oiii. 
The \nii\ images of four targets are displayed in 
Figure~\ref{F-images}. \object{IC~5117} and \object{NGC~6833} are well 
resolved by the HST. \object{IC~5117} presents a highly elongated and 
clumpy morphology, possibly with a `multipolar' shape 
(cf. e.g. He~2-47 and M~1-37, Sahai \cite{Sa00}). \object{NGC 6833} 
shows a bright nucleus from which two `boxy' lobes depart. 
 
\object{BoBn 1} and \object{M 1-71} are also resolved in the 
ground-based images. \object{BoBn 1} in the \nii\ image shows a small 
protrusion toward the South-East. \object{M 1-71} is larger, and 
appears to be composed of a bright `bulge' and a fainter disk-like 
structure extending approximately North-South. 
 
\object{DdDm 1} and \object{M1-5} are only marginally resolved in our 
ground-based images, and no details of their morphology can be 
appreciated. For this reason, they are not displayed 
in Figure~\ref{F-images}. \object{DdDm 1} does not show any sign of 
elongation, while \object{M1-5} is slightly elliptical, as clearly 
shown in \oiii\ and \hb\ HST images kindly made available by Matt 
Bobrowsky. 
 
In Table~\ref{obstable} we list the sizes of the nebulae, measured as 
the diameter at the 10\% level of the peak surface brightness, 
deconvolved by the instrumental resolution profile (seeing in the case 
of ground-based images). This method was introduced by Tylenda et 
al. (\cite{Ty03}) and proved to be a good diameter estimator for these 
barely resolved images. For M~1-5, we used the HST image of Matt 
Bobrowsky.  
 
\section{Analysis}  
\label {S-analysis}  
 
\subsection{Line fluxes and equivalent widths} 
 
Reduced spectra were transferred to the STARLINK spectrum analysis 
program DIPSO (Howarth et al. \cite{How96}) to measure emission-line 
fluxes and equivalent widths. Local continua were fit with a 
first-degree polynomial on both sides of the emission-lines and 
subtracted from the spectra. Line fluxes were measured by integrating 
the continuum subtracted emission-lines between initial and ending 
points of emission features. Integration was performed to ensure we 
were deriving a `measured' flux as opposed to a fitted flux (e.g., 
from a Gaussian fit). Uncertainities were assigned to each flux line 
measurement based upon that frame's exposure time and flux calibration 
sensitivity. 

Equivalent widths were measured with respect to 
local linear continua.  Errors for each equivalent width measurement 
were derived by DIPSO based upon scattering of linear continuum fit 
and signal-to-noise of each emission-line. 
 
Total absolute fluxes and equivalent widths listed in Tables~3 through 
8 were computed from the 
weighted average of individual measurements in the various 
spectra. Their respective errors, also listed in the Tables, were 
derived from the weighted standard deviation of the individual
measurements for each emission line.  
 
It should be noted that most of the errors listed in Tables 3-8 are 
`internal' errors which take into account the photon statistics of the 
observed PNe and standard stars, and the instrument and detector 
properties. In particular, this is the case when multiple observations 
for an object were all taken on a single night, and applies to the data 
of BoBn~1 (Table~\ref{bobtable}), DdDm~1 (Table~\ref{dddm1table}), 
M~1-5 (Table~\ref{m15table}), and M~1-71 (Table~\ref{m171table}). 
When data are obtained in different nights, as for some emission 
lines of NGC~6833 and IC~5117 (noted in Table~\ref{ngc6833table} and 
Table~\ref{ic5117table}), a better estimate of `external' systematic 
errors is obtained, for instance from the use of particular standard 
stars, or a non-perfect knowledge of the atmospheric extinction. 
Errors for multiple nights are larger and should be considered more 
realistic than for observations based on just a single night. 
 
Another possible source of errors for the total fluxes of our target 
PNe are slit losses due to the finite width of the adopted slit. 
These have been kept as small as possible by selecting compact PNe and 
adopting a slit width of 7\arcsec\ for the spectroscopic 
observations. Slit losses have been estimated taking into account the 
size of each nebula in different emission lines from the present 
images, seeing effects, slit width and its orientation with respect to 
the long axis of the resolved nebula, as well as centering and guiding 
errors.  Slit losses are small in the case of the smaller nebulae, 
namely BoBn1, DdDm 1, and NGC 6833. Considering that seeing during 
the observations was generally not worse than 1.5 arcsec, slit 
losses for these nebulae are estimated to be below 0.5\%. The effect 
is slightly higher for M~1-5, were slit losses are estimated to be 
around 1.0\%.  They are potentially larger for IC~5117, especially in 
\nii\ where the nebula is most extended. However, except in M1-71, our
slit loss was oriented along the major axis of its elongated 
morphology, and thus no light from the outer regions of the nebula has 
escaped; slit losses are estimated to be at the same level as the 
first three nebulae, namely below 0.5\%.  The only case in which slit 
losses are substantial (approximately 3\% in \nii\ and 2\%\ in \ha) is 
M~1-71, as the slit (oriented along the parallactic angle) was cutting 
the nebula through its short axis, a fact only partially compensated 
by the fact that the emission is highly centrally concentrated. 
 
Including all the different sources of uncertainties discussed above, 
our estimates of the errors for the total fluxes to be associated with 
each individual PN are presented in Table~\ref{errors}. They are 
obtained by summing quadratically the errors due to slit 
losses with the average of the errors listed in Table~3-8 for all 
lines in each individual PN, or for only the brightest ones ($H\beta$, 
$H\alpha$, 5007\rm {\AA}, 6583\rm {\AA}, and 6678\rm {\AA}). 
 
The conclusion is that narrowband images in bright emission lines can 
be calibrated photometrically using all the nebulae in our sample 
except for M~1-71 to a level of 1.4\% or better, depending on the 
actual `standard' PN adopted. At the same level, secular variations of 
total flux of nebulae can be monitored by means of future 
spectrophotometric observations. 
 
\begin{table} 
\caption[]{Estimated total errors for all emission lines fluxes, and 
for available bright lines, namely $H\beta$, $H\alpha$, 5007\rm 
{\AA}, 6583\rm {\AA}, and 6678\rm {\AA}.} 
\begin{flushleft} 
\begin{tabular}{lcc} 
\hline \hline 
Object       & Totlal error& Total Error     \\ 
Name         & (all lines)& (bright lines) \\ 
\object {Bobn 1}  & 2.3\%      & 1.1\%  \\ 
\object {DdDm 1}  & 3.8\%      & $<$1\% \\ 
\object {IC 5117} & 2.6\%      & 1.1\%  \\ 
\object {M 1-5}   & 2.3\%      & $<$1\% \\ 
\object {M 1-71}  & 3.1\%      & 2.8\%  \\ 
\object {NGC 6833}& 5.2\%      & $<$1\%  \\ 
\hline 
\end{tabular} 
\end{flushleft} 
\label{errors} 
\end{table}

\subsection{Radial velocities} 
 
Derived systemic velocities are presented in 
Table~\ref{obstable}. Five out of six of our radial velocity 
measurements may be compared to previous literature, and in particular 
with the large compilation in Durand et al. (\cite{Dur98}), which also 
includes the data quoted in A92.  \object{NGC 6833}, \object{M 1-5},
and \object{BoBn 1} are in good agreement with the velocity 
quoted in Durand et al. (\cite{Dur98}). \object{IC 5117} is a factor 
of 2$\sigma$ in disagreement, while the radial velocity 
measured more recently by Hyung et al. (\cite{Hyu01}) is consistent  
with our determination.  Our \object{M 1-71} radial velocity 
measurement is in good agreement with the older measurements quoted 
in A92, however, it does not match the new 
measurements of Durand et al. (\cite{Dur98}) and their adopted value. 
Our high radial velocity for \object{DdDm 1} is in excellent agreement
with that reported by Barker \& Cudworth \cite{Bar84}.

\subsection{Physical conditions and chemical abundances} 
 
The present spectra allow us to derive physical and chemical 
properties for four PNe: \object{DdDm~1}, \object{IC~5117}, 
\object{NGC~6833}, and \object{M~1-71}. 

The same type of analysis as in Corradi et al. (\cite{C97}) and Perinotto \& 
Corradi (\cite{P98}) was adopted.  The logarithmic extinction 
constants \cbeta\ were computed from the Balmer decrement.  
Electron temperatures T$_e$(\nii) and T$_e$(\oiii) were computed from 
the standard auroral to nebular line ratios.  As with electron 
density, in all nebulae where the \sii\ 673.1/671.6~nm line ratio 
could be determined, it approaches its high density limit, implying 
N$_e$$>$10000~cm$^{-3}$.  For IC~5117 and DdDm~1, an estimate of N$_e$ 
can be obtained from the \cliii~551.8~nm, 553.8~nm doublet. The 
\cbeta, and electron temperatures and densities are listed in Table~9. 
 
Ionic abundances relative to hydrogen are computed given the temperature
and densities, using the line fluxes relative to \hb~(Table~10).
Total abundances and their errors (Table~11) were computed from  
line intensities as described in Corradi et al. (\cite{C97}), using
the ionization correction factor (ICF) from Kingsburgh \& Barlow \cite{KB94}
(hereafter; KB94). The low
level of scatter in measurements for each emission-line flux did 
propagate into final uncertainties in the derived abundances of the 
order of 5 percent for helium and 25 percent for heavier 
elements.

\section{Discussion and Conclusions} 
\label{S-conclusions}  
 
The spectra obtained were used to infer some physical and  
chemical properties of the target nebulae. This analysis generally  
confirms and extends results obtained by other authors.  
 
\object{BoBn 1} and \object{DdDm 1} are known to belong to the 
Galactic halo, owing to their high Galactic latitude, highly peculiar 
systemic radial velocity, and subsolar oxygen abundance (see also 
Barker \& Cudworth \cite{Bar84}, Howard et al. \cite{Ho97}, and
Dinerstein et al. \cite{Din03}). While our flux ratios for \object{DdDm 1}
do not agree well with Dinerstein et al. \cite{Din03}, we observe
no systematic effects between the three line ratios that correspond
to our wavelength range.
 
An extensive spectroscopic study of IC5117 was presented by Hyung et 
al. (\cite{Hyu01}), who show that this nebula must be very young owing 
to its high gas densities, ranging from 40\,000~cm$^{-3}$ in the 
\cliii\ and \sii\ emitting region up to 100\,000~cm$^{-3}$ from  
line diagnostics involving $p^3$ electrons. The large density 
fluctuations found in this nebula are most likely related to the 
clumpy and highly asymmetrical morphology shown by HST images 
presented in this paper. 

We found rather high gas densities in \object{M 1-71}.
Parthasarathy et al. (\cite{PAS98}) classified the central star 
of \object{M 1-71} in the class of ``weak emission line stars'' 
(WELS) or [WC]-PG1159 central stars. Our spectra confirm the presence 
of C{\sc IV} 580.6~nm and 465.0 to 468.6~nm C{\sc IV}-He{\sc II} 
broad stellar features in emission, but not of any O{\sc V} and O{\sc 
VI} lines. 
  
To our knowledge, no abundance determinations are available for the  
nebula of \object{M 1-71} prior to the present study. Our analysis  
confirms that, like in the other WELS and [WC] central stars of PNe,  
in spite of the fact that the central star of \object{M 1-71} is  
helium and carbon rich, no peculiar gas abundances are found in the  
nebula compared to the general sample of Galactic PNe (cf. Corradi \&  
Schwarz \cite{CS95}).   

\object{NGC 6833} has a peculiar systemic radial velocity, 
deviating more than 50~\kms\ from the general Galactic rotation, (even 
assuming a large distance of 8-9 kpc for this object).  
The galactic coordinates and radial velocity measurements of 
\object{NGC 6833} coincide well with the outer arm structure observed
in HI within the range of 50$\fdg$$<$ \emph{l} $<$ 195 $\fdg$,
-5$\fdg$$<$ \emph{b} $<$35$\fdg$, and -175 km s$^{-1}$$<$ V$_{LSR}$
$<$-60 km s$^{-1}$ (Wakker \& van Woerden \cite{Wak91},
Dwarakanath et al. \cite{Dwa02}). Low oxygen, neon and nitrogen abundances 
(cf. also the data for the global sample of PNe summmarized in Corradi \& 
Schwarz \cite{CS95}), suggests that \object{NGC 6833} is a nebula 
belonging to a relatively old stellar population.
The electron density lower limit of 10\,000~cm$^{3}$ 
suggests that this nebula is likely to be in an early evolutionary 
stage. 

For NGC~6833, our chemical abundances are in good agreement
with those listed in Perinotto (\cite{P91}).  Abundances and T$_e$(\oiii) 
of \object{IC 5117}, \object{NGC 6833}, and \object{DdDm 1} are also 
in good agreement with Mal'kov (\cite{Ma98}),
with abundances of He and heavier elements comparable within $<$5\%, 
and effective temperatures deviating on average by $\sim$250~K. 

By comparing our abundance measurements with KB94 average abundance
values for a particular class of PNe, we see that \object{IC 5117} and
\object{M 1-71} fall within the range for non-type 1 PNe. Our abundance 
measurements for \object{NGC 6833} (except He) are all significantly
below the average KB94 value for non-type 1 Pne; O, N, and Ne abundances
are a factor of 3.8,2.5, and 3.6 below, respectively and Ar and S are even
further below. If \object{NGC 6833} is associated with the outer arm, then its
abundance measurements would be an indicator of the outer arm's metallicity,
which has been estimated to be lower than the inner disk (Digel er al. 
\cite{Dig90}). However, we cannot exclude the possibility that it belongs to the 
Galactic halo.

For four PNe, \object{IC 5117}, \object{M 1-5}, \object{M 1-71}, and 
\object{NGC 6833}, \hb~flux falls within one sigma of A92
reported \hb~fluxes. \object{BoBn 1} and 
\object{DdDm 1} are greater than three sigma from A92. 
\hb~measurements. This discrepancy between our measurements and A92 may be
due to underestimated uncertainities within the A92 catalogue, which has 
been recently noted in Ruffle et al. \cite{Ruf04}. However, if the flux of 
\object{BoBn 1} and \object{DdDm 1} accurately represent intrinsic variability,
then this calls for future monitoring of these sources.
  
In conclusion, following the work done in the southern hemisphere by 
Dopita \& Hua (\cite{Dop97}), the present data provide an extension to 
the northern hemisphere of new emission-line standards for narrowband 
imaging. Further work of this kind is needed in order to have more 
complete coverage in right ascension and radial velocities. 
  
\begin{acknowledgements}  
  
The authors thank Bego\~na Garc\'\i a for one night of service  
observations, Denise Gon\c calves for further observations, Danny  
Lennon for his thoughtful help in data reduction and analysis, 
Remington P.S. Stone for valuable comments, and the referee for a
number of helpful insights.
  
\end{acknowledgements}  
  
%

\begin{table*}  
\caption[]{\object{BoBn 1} fluxes and equivalent widths.}  
\begin{flushleft}  
\centering  
\begin{tabular}{lcccc}  
\hline \hline  
Wavelength &   Ion  &       log [Flux]        &      log [EW]       & Number of\\  
(\rm {\AA})&        &(ergs cm$^{-2}$ s$^{-1}$)&     (\rm {\AA})     & Observations\\  
\hline  
4685.71  & HeII      & $-13.103\pm0.024$    & $2.097\pm0.107$ & 3 \\   
4861.29  & $H\beta$  & $-12.425\pm0.011$    & $2.759\pm0.128$ & 4 \\   
4959.52  & [OIII]    & $-12.358\pm0.005$    & $2.797\pm0.065$ & 4 \\   
5007.57  & [OIII]    & $-11.877\pm0.006$    & $3.188\pm0.140$ & 4 \\   
5754.59  & [NII]     & $-14.447\pm0.045$    & $1.000\pm0.105$ & 2 \\   
5875.97  & HeI       & $-13.278\pm0.019$    & $2.009\pm0.255$ & 4 \\   
6300.30  & [OI]      & $-14.503\pm0.069$    & $0.986\pm0.050$ & 2 \\   
6548.05  & [NII]     & $-13.296\pm0.014$    & $1.674\pm0.045$ & 4 \\   
6562.85  & $H\alpha$ & $-11.934\pm0.002$    & $2.959\pm0.059$ & 4 \\   
6583.45  & [NII]     & $-12.828\pm0.012$    & $2.166\pm0.067$ & 4 \\   
6678.15  & HeI	     & $-13.853\pm0.022$    & $1.585\pm0.206$ & 3 \\   
7065.71  & HeI	     & $-13.723\pm0.038$    & $1.767\pm0.109$ & 3 \\   
\hline  
\end{tabular}  
\end{flushleft}  
\label{bobtable}  
\end{table*}  
  
\begin{table*}  
\centering  
\caption []{\object{DdDm 1} fluxes and equivalent widths}  
\begin{flushleft}  
\centering  
\begin{tabular}{lcccc}   
\hline \hline  
Wavelength &   Ion  &       log [Flux]        &      log [EW]      & Number of   
\\  
(\rm {\AA})&	    &(ergs cm$^{-2}$ s$^{-1}$)&     (\rm {\AA})    & Observations   
\\  
\hline  
3726.19 & [OII]    & $-11.759\pm0.001$  & $2.184\pm0.082$ & 2 \\  
3750.15 & $H_{12}$ & $-13.321\pm0.055$  & $0.719\pm0.045$ & 2 \\   
3770.63 & $H_{11}$ & $-13.257\pm0.043$  & $0.805\pm0.040$ & 3 \\   
3797.90 & $H_{10}$ & $-13.119\pm0.035$  & $0.938\pm0.052$ & 2 \\   
3819.7  & HeI*     & $-13.735\pm0.120$  & $0.267\pm0.102$ & 3 \\   
3835.38 & $H_{9}$  & $-12.969\pm0.031$  & $1.077\pm0.024$ & 4 \\   
3868.71 & [NeIII]  & $-12.313\pm0.009$  & $1.752\pm0.030$ & 3 \\   
3889.05 & $H_{8}$  & $-12.511\pm0.003$  & $1.562\pm0.068$ & 3 \\   
3967.41 & [NeIII]  & $-12.382\pm0.006$  & $1.724\pm0.015$ & 2 \\   
4026.1  & HeI*     & $-13.466\pm0.011$  & $0.672\pm0.033$ & 2 \\   
4068.91 & CIII     & $-13.530\pm0.024$  & $0.611\pm0.023$ & 4 \\   
4101.74 & $H\delta$& $-12.400\pm0.017$  & $1.742\pm0.041$ & 4 \\   
4340.47 & $H\gamma$& $-12.124\pm0.010$  & $2.096\pm0.034$ & 3 \\   
4363.21 & [OIII]   & $-13.096\pm0.012$  & $1.126\pm0.121$ & 2 \\   
4387.93 & HeI      & $-14.101\pm0.163$  & $0.135\pm0.253$ & 3 \\   
4471.68 & HeI      & $-13.098\pm0.021$  & $1.158\pm0.066$ & 3 \\   
4861.20 & $H\beta$ & $-11.794\pm0.001$  & $2.542\pm0.032$ & 3 \\   
4959.52 & [OIII]   & $-11.623\pm0.004$  & $2.699\pm0.036$ & 2 \\   
5007.57 & [OIII]   & $-11.145\pm0.0001$ & $3.034\pm0.120$ & 2 \\   
5197.90 & [NI]     & $-14.144\pm0.002$  & $0.298\pm0.011$ & 2 \\   
5269.20 & [KVI]    & $-13.714\pm0.019$  & $0.767\pm0.002$ & 2 \\   
5517.72 & [CIIII]  & $-14.459\pm0.132$  & $0.065\pm0.096$ & 2 \\   
5537.89 & [CIIII]  & $-14.339\pm0.158$  & $0.173\pm0.155$ & 2 \\   
5754.59 & [NII]    & $-13.683\pm0.025$  & $0.906\pm0.040$ & 2 \\  
\hline  
\end{tabular}  
\end{flushleft}  
\label{dddm1table}  
\centering  
$^\star$HeI lines are a double blend  
\end{table*}

\begin{table*}  
\caption[]{\object{IC 5117} fluxes and equivalent widths.}  
\begin{flushleft}  
\centering  
\begin{tabular}{llllc}  
\hline \hline  
Wavelength &   Ion  &       log [Flux]        &      log [EW]       & Number of\\  
(\rm {\AA})&        &(ergs cm$^{-2}$ s$^{-1}$)&     (\rm {\AA})     & Observations\\  
\hline  
3705.00 & HeI      & $-13.661\pm0.155$ & $0.810\pm0.211$ & 2 \\   
3712.75 & HeII     & $-13.656\pm0.032$ & $0.806\pm0.122$ & 2 \\   
3726.19 & [OII]    & $-12.337\pm0.004$ & $1.768\pm0.023$ & 2 \\   
3734.37 & $H_{13}$ & $-13.340\pm0.088$ & $0.493\pm0.261$ & 2 \\   
3750.15 & $H_{12}$ & $-13.254\pm0.021$ & $1.217\pm0.179$ & 2 \\   
3770.63 & $H_{11}$ & $-13.156\pm0.053$ & $1.502\pm0.051$ & 2 \\   
3797.90 & $H_{10}$ & $-13.014\pm0.005$ & $1.651\pm0.175$ & 2 \\   
3819.70 & HeI*     & $-13.700\pm0.001$ & $0.846\pm0.019$ & 2 \\   
3835.38 & $H_{9}$  & $-12.836\pm0.015$ & $1.767\pm0.016$ & 2 \\   
3868.71 & [NeIII]  & $-11.503\pm0.002$ & $2.871\pm0.071$ & 4 \\   
3889.05 & $H_{8}$  & $-12.482\pm0.013$ & $1.955\pm0.098$ & 3 \\   
3967.41 & [NeIII]  & $-11.844\pm0.003$ & $2.781\pm0.100$ & 4 \\   
4068.91 & CIII     & $-12.991\pm0.002$ & $1.382\pm0.063$ & 2 \\   
4101.74 & $H\delta$& $-12.179\pm0.012$ & $2.387\pm0.134$ & 4 \\   
4340.47 & $H\gamma$& $-11.876\pm0.003$ & $2.581\pm0.112$ & 4 \\   
4363.21 & [OIII]   & $-12.191\pm0.010$ & $2.262\pm0.077$ & 4 \\   
4387.93 & HeI      & $-13.878\pm0.152$ & $0.637\pm0.111$ & 2 \\   
4471.68 & HeI      & $-12.798\pm0.001$ & $1.783\pm0.024$ & 2 \\   
4634.14 & NIII     & $-13.576\pm0.036$ & $0.695\pm0.333$ & 4 \\   
4640.64 & NIII     & $-13.215\pm0.013$ & $0.947\pm0.291$ & 4 \\   
4685.71 & HeII     & $-12.359\pm0.001$ & $2.106\pm0.075$ & 4 \\   
4713.38 & HeI*     & $-13.036\pm0.003$ & $1.419\pm0.082$ & 4 \\   
4724.30 & [NeIV]   & $-14.154\pm0.057$ & $0.270\pm0.163$ & 4 \\   
4740.18 & [ArIV]   & $-12.661\pm0.001$ & $1.813\pm0.029$ & 4 \\   
4861.20 & $H\beta$ & $-11.380\pm0.020$ & $3.032\pm0.006$ & 6 \\   
4921.93 & HeI      & $-13.180\pm0.050$ & $1.019\pm0.063$ & 4 \\   
4958.52 & [OIII]   & $-10.629\pm0.023$ & $3.270\pm0.124$ & 5 \\   
5007.57 & [OIII]   & $-10.131\pm0.017$ & $3.743\pm0.026$ & 2 \\   
5191.80 & [ArIII]  & $-14.152\pm0.159$ & $0.151\pm0.004$ & 4 \\   
5197.90 & [NI]     & $-13.852\pm0.019$ & $0.488\pm0.116$ & 4 \\   
5411.52 & HeII     & $-13.276\pm0.042$ & $1.107\pm0.019$ & 4 \\   
5517.72 & [CIIII]  & $-14.084\pm0.025$ & $0.209\pm0.072$ & 4 \\   
5537.89 & [CIIII]  & $-13.544\pm0.005$ & $0.801\pm0.007$ & 4 \\   
5754.59 & [NII]    & $-12.653\pm0.006$ & $1.650\pm0.005$ & 2 \\   
5875.97 & HeI      & $-11.909\pm0.024$ & $2.325\pm0.109$ & 9 $\dagger$ \\   
6300.30 & [OI]     & $-12.177\pm0.026$ & $1.881\pm0.050$ & 8 $\dagger$ \\   
6312.10 & [SIII]   & $-12.606\pm0.039$ & $1.437\pm0.011$ & 9 $\dagger$ \\   
6363.77 & [OI]     & $-12.656\pm0.024$ & $1.453\pm0.038$ & 9 $\dagger$ \\   
6548.05 & [NII]    & $-11.733\pm0.032$ & $1.612\pm0.116$ & 9 $\dagger$ \\   
6562.85 & $H\alpha$& $-10.524\pm0.002$ & $2.887\pm0.223$ & 3 $\dagger$ \\   
6583.45 & [NII]    & $-11.272\pm0.008$ & $2.312\pm0.184$ & 7 $\dagger$ \\   
6678.15 & HeI      & $-12.366\pm0.003$ & $1.743\pm0.007$ & 9 $\dagger$ \\   
6716.47 & [SII]    & $-12.958\pm0.006$ & $1.151\pm0.015$ & 9 $\dagger$ \\   
6730.85 & [SII]    & $-12.612\pm0.008$ & $1.495\pm0.037$ & 9 $\dagger$ \\   
7065.71 & HeI      & $-11.860\pm0.009$ & $2.200\pm0.044$ & 9 $\dagger$ \\   
\hline  
\end{tabular}  
\end{flushleft}  
\label{ic5117table}  
$^\star$HeI lines are a double blend.\\  
$\dagger$ Measured values were determined from two nights of observation  
\end{table*}  
  
\begin{table*}  
\caption[]{\object{M 1-5} fluxes and equivalent widths.}  
\begin{flushleft}  
\centering  
\begin{tabular}{lcccc}  
\hline \hline  
Wavelength &   Ion  &       log [Flux]        &      log [EW]       & Number of\\  
(\rm {\AA})&        &(ergs cm$^{-2}$ s$^{-1}$)&     (\rm {\AA})     & Observations\\  
\hline  
5754.59 & [NII]     & $-13.355\pm0.098$ & $1.228\pm0.102$ & 4 \\   
5875.97 & HeI       & $-12.551\pm0.007$ & $2.029\pm0.062$ & 4 \\   
6300.30 & [OI]      & $-13.273\pm0.032$ & $1.265\pm0.038$ & 4 \\   
6312.10 & [SIII]    & $-13.685\pm0.034$ & $0.819\pm0.048$ & 3 \\   
6363.77 & [OI]      & $-13.738\pm0.080$ & $0.779\pm0.060$ & 3 \\   
6548.05 & [NII]     & $-12.086\pm0.002$ & $1.793\pm0.076$ & 4 \\   
6562.85 & $H\alpha$ & $-11.082\pm0.002$ & $2.817\pm0.029$ & 2 \\   
6583.45 & [NII]     & $-11.595\pm0.001$ & $2.498\pm0.063$ & 4 \\   
6678.15 & HeI       & $-12.940\pm0.001$ & $1.572\pm0.032$ & 2 \\   
6716.47 & [SII]     & $-13.497\pm0.005$ & $1.008\pm0.026$ & 2 \\  
6730.85 & [SII]     & $-13.157\pm0.001$ & $1.332\pm0.016$ & 2 \\  
7065.71 & HeI       & $-12.513\pm0.002$ & $1.942\pm0.031$ & 4 \\  
7135.80 & [ArIII]   & $-12.530\pm0.010$ & $1.919\pm0.045$ & 4 \\  
\hline  
\end{tabular}  
\end{flushleft}  
\label{m15table}  
\end{table*}  
  
\begin{table*}  
\caption[]{\object{M 1-71} fluxes and equivalent widths. Note that an 
additional uncertainty of some 2-3\% due to slit losses should be 
added to the errors quoted below (see text).} 
\begin{flushleft}  
\centering  
\begin{tabular}{lcccc}  
\hline \hline  
Wavelength &   Ion  &       log [Flux]        &      log [EW]      & Number of\\  
(\rm {\AA})&        &(ergs cm$^{-2}$ s$^{-1}$)&     (\rm {\AA})    & Observations\\  
\hline  
3868.71 & [NeIII]   & $-12.683\pm0.032$ & $2.413\pm0.163$ & 2 \\   
3967.41 & [NeIII]   & $-12.968\pm0.053$ & $2.295\pm0.161$ & 2 \\   
4101.74 & $H\delta$ & $-13.188\pm0.007$ & $2.027\pm0.158$ & 2 \\   
4340.47 & $H\gamma$ & $-12.803\pm0.036$ & $2.420\pm0.073$ & 2 \\   
4363.21 & [OIII]    & $-13.554\pm0.019$ & $1.552\pm0.089$ & 2 \\   
4471.68 & HeI 	    & $-13.615\pm0.039$ & $1.744\pm0.041$ & 2 \\   
4861.20 & $H\beta$  & $-12.198\pm0.048$ & $2.911\pm0.228$ & 4 \\   
4959.52 & [OIII]    & $-11.510\pm0.010$ & $3.207\pm0.045$ & 3 \\   
5007.57 & [OIII]    & $-10.997\pm0.002$ & $3.645\pm0.147$ & 2 \\   
5754.59 & [NII]     & $-13.210\pm0.011$ & $1.677\pm0.017$ & 2 \\   
5875.97 & HeI       & $-12.409\pm0.001$ & $2.394\pm0.061$ & 2 \\   
6300.30 & [OI]      & $-12.625\pm0.007$ & $1.888\pm0.104$ & 2 \\   
6312.10 & [SIII]    & $-13.597\pm0.001$ & $0.759\pm0.229$ & 2 \\   
6363.77 & [OI]      & $-13.097\pm0.034$ & $1.598\pm0.050$ & 2 \\   
6548.05 & [NII]     & $-11.961\pm0.011$ & $1.541\pm0.069$ & 2 \\   
6562.85 & $H\alpha$ & $-10.982\pm0.001$ & $2.590\pm0.052$ & 1 $\ddagger$ \\   
6583.45 & [NII]     & $-11.470\pm0.006$ & $2.469\pm0.035$ & 2 \\   
6678.15 & HeI       & $-12.751\pm0.006$ & $1.802\pm0.004$ & 2 \\   
6716.47 & [SII]     & $-13.334\pm0.029$ & $1.067\pm0.093$ & 2 \\   
6730.85 & [SII]     & $-13.027\pm0.012$ & $1.462\pm0.024$ & 2 \\  
7065.71 & HeI       & $-12.282\pm0.008$ & $2.193\pm0.035$ & 2 \\  
\hline  
\end{tabular}  
\end{flushleft}  
\label{m171table}  
$\ddagger$ Errors are based on one measurement. \\  
\end{table*}  
  
\begin{table*}  
\caption[]{\object{NGC 6833} fluxes and equivalent widths.}  
\begin{flushleft}  
\centering  
\begin{tabular}{lcccc}   
\hline \hline  
Wavelength &   Ion  &       log [Flux]        &      log [EW]      & Number of\\  
(\rm {\AA})&        &(ergs cm$^{-2}$ s$^{-1}$)&     (\rm {\AA})    & Observations\\  
\hline  
3697.15 & $H_{17}$ & $-13.829\pm0.241$ & $-0.0003\pm0.109$& 4 $\dagger$ \\   
3705.00 & HeI      & $-13.288\pm0.081$ & $0.423\pm0.037$  & 4 $\dagger$ \\   
3712.75 & HeII     & $-13.346\pm0.149$ & $0.401\pm0.043$  & 4 $\dagger$ \\   
3726.19 & [OII]    & $-12.259\pm0.066$ & $1.358\pm0.012$  & 4 $\dagger$ \\   
3734.37 & $H_{13}$ & $-12.977\pm0.116$ & $0.562\pm0.024$  & 4 $\dagger$ \\   
3750.15 & $H_{12}$ & $-12.920\pm0.023$ & $0.959\pm0.004$  & 6 $\dagger$ \\   
3770.63 & $H_{11}$ & $-12.812\pm0.024$ & $1.066\pm0.005$  & 6 $\dagger$ \\   
3797.90 & $H_{10}$ & $-12.689\pm0.055$ & $1.184\pm0.003$  & 6 $\dagger$ \\   
3819.70 & HeI*     & $-13.305\pm0.146$ & $0.573\pm0.076$  & 5 $\dagger$ \\   
3835.38 & $H_{9}$  & $-12.540\pm0.030$ & $1.334\pm0.006$  & 6 $\dagger$ \\   
3868.71 & [NeIII]  & $-11.439\pm0.020$ & $2.394\pm0.033$  & 8 $\dagger$ \\   
3889.05 & $H_{8}$  & $-12.142\pm0.019$ & $1.692\pm0.024$  & 7 $\dagger$\\   
3967.41 & [NeIII]  & $-11.761\pm0.001$ & $2.136\pm0.005$  & 8 $\dagger$ \\   
4026.10 & HeI*     & $-13.025\pm0.013$ & $0.905\pm0.014$  & 5 $\dagger$ \\   
4068.91 & CIII     & $-13.442\pm0.128$ & $0.481\pm0.117$  & 4 $\dagger$ \\   
4101.74 & $H\delta$& $-11.954\pm0.006$ & $1.988\pm0.001$  & 7 $\dagger$ \\   
4340.47 & $H\gamma$& $-11.673\pm0.037$ & $2.269\pm0.026$  & 8 $\dagger$ \\   
4363.21 & [OIII]   & $-12.247\pm0.040$ & $1.715\pm0.056$  & 8 $\dagger$ \\   
4387.93 & HeI      & $-13.601\pm0.042$ & $0.378\pm0.006$  & 5 $\dagger$ \\   
4471.68 & HeI      & $-12.613\pm0.039$ & $1.369\pm0.030$  & 7 $\dagger$ \\   
4713.38 & HeI      & $-13.244\pm0.009$ & $0.821\pm0.012$  & 6 $\dagger$ \\   
4740.18 & [ArIV]   & $-13.447\pm0.034$ & $0.590\pm0.039$  & 5 $\dagger$ \\   
4861.20 & $H\beta$ & $-11.315\pm0.012$ & $2.733\pm0.033$  & 7 $\dagger$ \\   
4921.93 & HeI      & $-13.246\pm0.034$ & $0.785\pm0.050$  & 4 $\dagger$ \\   
4959.52 & [OIII]   & $-10.886\pm0.007$ & $3.003\pm0.023$  & 4 $\dagger$ \\   
5007.57 & [OIII]   & $-10.401\pm0.001$ & $3.439\pm0.074$  & 2 $\dagger$ \\   
5191.80 & [ArIII]  & $-14.275\pm0.177$ & $-0.150\pm0.186$ & 2 \\   
5537.89 & [CIIII]  & $-13.935\pm0.082$ & $0.243\pm0.098$  & 2 \\   
5754.59 & [NII]    & $-13.049\pm0.047$ & $1.121\pm0.017$  & 6 \\   
5875.97 & HeI      & $-12.038\pm0.002$ & $2.112\pm0.011$  & 3 \\   
6300.30 & [OI]     & $-12.841\pm0.268$ & $1.342\pm0.272$  & 3 \\   
6363.77 & [OI]     & $-13.245\pm0.006$ & $0.966\pm0.080$  & 3 \\   
6548.05 & [NII]    & $-12.262\pm0.007$ & $1.247\pm0.038$  & 4 \\   
6562.85 & $H\alpha$& $-10.752\pm0.001$ & $2.838\pm0.051$  & 1 $\ddagger$ \\   
6583.45 & [NII]    & $-11.865\pm0.003$ & $1.867\pm0.021$  & 4 \\   
6678.15 & HeI      & $-12.620\pm0.007$ & $1.598\pm0.028$  & 4 \\   
6716.47 & [SII]    & $-13.888\pm0.033$ & $0.329\pm0.035$  & 3 \\   
6730.85 & [SII]    & $-13.531\pm0.029$ & $0.700\pm0.001$ & 3 \\   
7065.71 & HeI      & $-12.130\pm0.003$ & $2.123\pm0.009$  & 3 \\   
\hline  
\end{tabular}  
\end{flushleft}  
\label{ngc6833table}  
$^\star$HeI lines are a double blend.\\  
$\dagger$ Measured values were determined from two nights of observation. \\  
$\ddagger$ Errors are based on one measurement.  
\end{table*}  
 
\begin{table*}  
\begin{flushleft}  
\caption[]{Logarithmic extinction constant \cbeta, and electron temperatures  
and densities. Errors on temperatures are within 500 Kelvin.}  
\begin{tabular}{lccccc}   
\hline \hline  
Object &\cbeta\ & T$_e$(\oiii) & T$_e$(\nii) & N$_e$(\sii)  & N$_e$(\cliii)     \\  
Name   &        &    (K)    &   (K)    & (cm$^{-3}$)  & (cm$^{-3}$)            \\   
\hline	        		             
DdDm1  &   -    & 12\,200    &    -      &     -    &    3\,700    \\    
IC5117 &  1.30  & 12\,500    & 13\,500   &     -    &   38\,000    \\  
NGC6833&  0.34  & 13\,200    & 15\,400   & $>$10\,000 &      -       \\  
M1-71  &  2.44  &  9\,800    & 10\,200   & $>$10\,000 &      -      \\  
\hline  
\end{tabular}  
\end{flushleft}  
\label{physical}  
\end{table*}  
  
\begin{table*}  
\begin{flushleft}  
\caption[]{Ionic abundances relative to hydrogen.}  
\begin{tabular}{lccccccccccc} \hline \hline  
Object &   HeI   &  HeII   &   OI    &   OII   &  OIII   &   NI    &   NII   &   SII   &  SIII   &  NeIII  &  ArIV   
\\  
\hline  
DdDm 1  & 1.04e-1 &    -   &    -   &    -   & 8.6e-5 &    -   &    -   &    -   &    -   & 1.5e-5 &    -  
\\  
IC 5117 & 9.33e-2 & 1.0e-2 & 5.4e-6 & 1.3e-5 & 2.9e-4 & 1.3e-6 & 5.9e-6 & 1.8e-7 & 2.0e-6 &    -   &    -  
\\  
M 1-71  & 1.17e-1 &    -   & 1.6e-5 &    -   & 5.0e-4 &    -   & 2.1e-5 & 3.4e-7 & 1.8e-6 & 1.5e-4 &    -  
\\  
NGC 6833& 9.05e-2 &    -   & 1.3e-6 & 4.5e-6 & 1.2e-4 &    -   & 2.0e-6 & 3.0e-8 &    -   & 3.4e-5 & 4.7e-8  
\\  
\hline  
\end{tabular}  
\end{flushleft}  
\label{T-ionabund}  
\end{table*}  
  
\begin{table*}  
\begin{flushleft}  
\caption[]{Total abundances and ionization correction factors (ICF).}  
\begin{tabular}{lccccccccccc}   
\hline \hline  
Object &    He   &    O    & ICF[O] &    N    & ICF[N] &    S    & ICF[S] &   Ne    & ICF[Ne] &   Ar    & ICF[Ar]  
\\  
\hline  
IC 5117 & 1.04e-1 & 3.2e-4 & 1.07  & 1.5e-4 & 25.51 & 4.4e-6 & 2.07  &    -   &    -   &    -   &   -  
\\  
M 1-71  & 1.17e-1 & 5.0e-4 & 1.00  &    -   &   -   &    -   &   -   & 1.5e-4 &  1.00  &    -   &   -  
\\  
NGC 6833& 9.05e-2 & 1.3e-4 & 1.00  & 5.6e-5 & 27.98 & 6.3e-8 & 2.13  & 3.5e-5 &  1.04  & 4.9e-8 & 1.04  
\\  
\hline  
\end{tabular}  
\end{flushleft}  
\label{T-totabund}  
\end{table*}  
  
\end{document}